# Reducing the impact of non-ideal PRBS on microwave photonic random demodulators by low biasing the optical modulator via PRBS amplitude compression

Shiyang Liu [a,b], Yang Chen[a,b,*]

[a] Shanghai Key Laboratory of Multidimensional Information Processing, East China Normal University, Shanghai, 200241, China
[b] Engineering Center of SHMEC for Space Information and GNSS, East China Normal University, Shanghai, 200241, China
[*] ychen@ce.ecnu.edu.cn

**ABSTRACT**
A novel method for reducing the impact of non-ideal pseudo-random binary sequence (PRBS) on microwave photonic random demodulators (RDs) in a photonics-assisted compressed sensing (CS) system is proposed. Different from the commonly used method that switches the bias point of the optical modulator in the RD between two quadrature transmission points to mix the signal to be sampled and the PRBS, this method employs a PRBS with lower amplitude to low bias the optical modulator so that the impact of non-ideal PRBS on microwave photonic RDs can be greatly reduced by compressing the amplitude of non-ideal parts of the PRBS. An experiment is performed to verify the concept. The optical modulator is properly low-biased via PRBS amplitude compression. The data rate and occupied bandwidth of the PRBS are 500 Mb/s and 1 GHz, while the multi-tone signals with a maximum frequency of 100 MHz are sampled at an equivalent sampling rate of only 50 MSa/s. The results show that the reconstruction error can be reduced by up to 85%. The proposed method can significantly reduce the requirements for PRBS in RD-based photonics-assisted CS systems, providing a feasible solution for reducing the complexity and cost of system implementation.

## 1. Introduction
Due to its potential advantages in the generation, processing, and transmission of high-frequency and large-bandwidth electrical signals, microwave photonics has been experiencing increasing development in the past few decades [1]. Among the many achievable functions, analog-to-digital conversion (ADC) is one important function [2]. According to the Nyquist sampling theorem, the ADC needs to have a sampling rate twice the maximum frequency of the signal to be sampled, which results in high cost and complexity when sampling high-frequency signals. Compressed sensing (CS) takes a different path and provides a method to receive signals at a lower sampling rate [3]. Due to the huge advantages brought by microwave photonics, photonics-assisted CS technology has received more and more attention in recent years [4].

Currently, the two most prevalent CS systems are the random demodulator (RD) and the modulated wideband converter. The RD, known for its easy implementation and simple structure, has attracted significant research interest from the photonic society, and numerous studies have been conducted [5-8]. In a microwave photonic RD, the signal to be sampled is first mixed with a pseudo-random binary sequence (PRBS) in a microwave photonic mixer, and then the mixed signal will be low-pass filtered and downsampled by a low-speed ADC. Finally, the original signal can be reconstructed by the sparse reconstruction algorithm digitally. According to the CS theory, the signal to be sampled needs to be multiplied by the PRBS, so its waveform is randomly multiplied by +1 or -1 at different times. Thus, in the mixing stage, the PRBS needs to have an ideal waveform theoretically, and the non-ideality of the PRBS, such as the waveform distortion and the amplitude jitters will seriously affect the reconstruction quality of the signal [9].

However, the RD requires that the bit rate of the PRBS must be greater than the Nyquist rate of the signal to be sampled. When sampling high-frequency signals, a very high-speed PRBS that is close to ideal is required. As it occupies a very large bandwidth, it is difficult to generate the close to ideal PRBS. Refs. [9, 10] have pointed out that the non-ideality of the PRBS can seriously affect the quality of the reconstructed signal, and the problem of non-ideal PRBS has been quantified and analyzed in detail in Ref. [9]. Nevertheless, few works focus on this issue in photonics-assisted CS systems. In fact, due to the significant differences between the optical and electrical RD structures, this impact is even more significant in microwave photonic RD. In addition, most of the relevant works have only analyzed the impact of the non-ideal PRBS brought to the CS systems, but have not proposed a solution to eliminate or mitigate this effect.

To solve the problem in microwave photonic RDs, in this letter, a novel method for reducing the impact of non-ideal PRBS on microwave photonic RDs is proposed. Different from the commonly used method that switches the optical modulator in the RD between two quadrature transmission points (QTPs) to mix the signal to be sampled and the PRBS, in the proposed method, we employ a PRBS with lower amplitude to low bias the optical modulator so that the impact of the non-ideal PRBS on microwave photonic RDs can be greatly reduced by compressing the amplitude of non-ideal parts of the PRBS. In moving the bias point from the QTP to the minimum transmission point (MITP), although the gain of the signal will decrease, the second harmonic interference will appear and increase, and the signal-to-noise ratio (SNR) will deteriorate when the relative intensity noise is not dominant, the performance improvement brought by compressing the amplitude of the non-ideal part of the PRBS is experimentally verified to be dominant in a certain degree of low bias. An experiment is performed to verify the concept. When the optical modulator is properly low-biased, the data rate and occupied bandwidth of the PRBS are 500 Mb/s and 1 GHz, and the multi-tone signals with a maximum frequency of 100 MHz are sampled at an equivalent sampling rate of 50 MSa/s, the reconstruction error can be reduced by up to 85%.

## 2. Principle

The schematic of the microwave photonic RD is shown in Fig. 1. A laser diode (LD, HLT-ITLA-M-C20-1-1-FA) generates an optical signal that is used as an optical carrier and injected into the dual-drive Mach–Zehnder modulator (DD-MZM, Fujitsu FTM7937EZ200) via an optical isolator and a polarization controller. The DD-MZM is biased and stabilized at the MITP using a bias controller. The non-ideal PRBS and the signal to be sampled are simultaneously generated using a four-port arbitrary waveform generator (AWG, Keysight M8195A) and applied to the two RF ports of the DD-MZM after being amplified by two electrical amplifiers (EAs, Multilink MTC5515). The optical signal from the DD-MZM is injected into a photodetector (PD, Bookham PP-10G) to finally mix the signal to be sampled and the PRBS. The mixed electrical signal from the PD is sampled by the analog-to-digital converter (ADC) in an oscilloscope (OSC, R&S RTO2032). The low-pass filtering and downsampling are carried out in the digital domain after sampling. Subsequently, the original signal can be reconstructed from the downsampled signal by the sparse reconstruction algorithm. In this work, the sparse reconstruction algorithm is the orthogonal matching pursuit algorithm. Because the signal to be sampled in this work is the multi-tone signal, the sparse basis used in this work is the Fourier orthogonal basis.

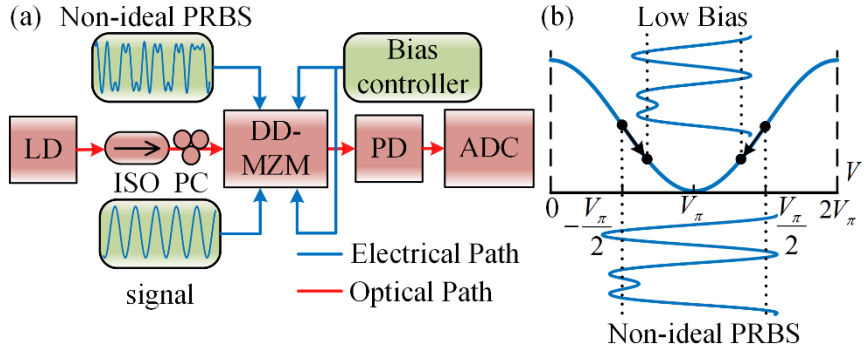

Fig. 1. (a) Schematic diagram of the microwave photonic RD. (b) Concept of low-biasing the DD-MZM via PRBS amplitude compression. LD, laser diode; ISO, optical isolator; PC, polarization controller; DD-MZM, dual-drive Mach–Zehnder modulator; PD, photodetector; ADC, analog-to-digital converter.

In the above process, the normalized output of the DD-MZM can be expressed as:

$$E_{out}(t) = \frac{1}{2} + \frac{1}{2}\cos\left(\frac{\pi}{V_\pi}\Delta V\right), \quad (1)$$

where $V_\pi$ is the half-wave voltage of the DD-MZM, and $\Delta V$ is the voltage difference between the upper and lower arms of the DD-MZM, including both the RF voltage and DC voltage.

Assuming that the static bias voltage is equal to $V_\pi$, when PRBS switches between $V_\pi/2$ and $-V_\pi/2$, it can be considered that the actual bias point of the DD-MZM switches between the positive QTP and the negative QTP as shown in Fig. 1(b). When the bias voltage is equal to $V_\pi$, Eq. (2) can be further written as:

$$\begin{aligned}
E_{out}(t) &= \frac{1}{2} + \frac{1}{2}\cos\left[\frac{\pi}{V_\pi}\left(V_{DC} + V_{PRBS}(t) + V_x(t)\right)\right] \\
&= \frac{1}{2} + \frac{1}{2}\cos\left[\frac{\pi}{V_\pi}\left(V_\pi + V_{PRBS}(t)\right)\right]\cos\left[V_x(t)\right] \\
&\quad - \frac{1}{2}\sin\left[\frac{\pi}{V_\pi}\left(V_\pi + V_{PRBS}(t)\right)\right]\sin\left[V_x(t)\right] \\
&\approx \frac{1}{2} - \frac{1}{2}\cos\left[\frac{\pi}{V_\pi}V_{PRBS}(t)\right]\left[1 - \frac{1}{2}V_x^2(t)\right] + \frac{1}{2}\sin\left[\frac{\pi}{V_\pi}V_{PRBS}(t)\right]V_x(t), \quad (2)\\
&= \frac{1}{2}\sin\left[\frac{\pi}{V_\pi}V_{PRBS}(t)\right]V_x(t) + \frac{1}{4}\cos\left[\frac{\pi}{V_\pi}V_{PRBS}(t)\right]V_x^2(t) \\
&\quad - \frac{1}{2}\cos\left[\frac{\pi}{V_\pi}V_{PRBS}(t)\right] + \frac{1}{2}
\end{aligned}$$

where $V_{PRBS}(t) = V_{code}s(t)$ is the PRBS waveform, $V_{code}$ is the amplitude of the PRBS waveform, $s(t)$ is the PRBS sequence, and $V_x(t)$ is the signal to be sampled. In the above derivation, the small signal modulation condition is used ($V_x(t) \ll 1$).

When an ideal PRBS with an amplitude of $V_{code} = V_\pi/2$ is used, Eq. (2) can be further simplified as

$$E_{out}(t) = \frac{1}{2}s(t)V_x(t) + \frac{1}{2} = \begin{cases} \frac{1}{2}V_x(t) + \frac{1}{2} & s(t) = 1 \\ -\frac{1}{2}V_x(t) + \frac{1}{2} & s(t) = -1 \end{cases}, \quad (3)$$

When an AC-coupled PD is used, the DC term can be ignored, and it can be seen from Eq. (3) that the mixing of the PRBS and the signal to be sampled is realized using an ideal PRBS [11].

However, when the PRBS does not have an ideal waveform, the above results in Eq. (3) cannot be obtained. For example, when the bandwidth of the PRBS is limited, there will be slow rise/fall edges between −1 and +1 code, and at the same time, the waveform will exhibit jitters around $\pm V_\pi/2$ during the original -1 and +1 code duration. In this case, Eq. (2) can be written as

$$E_{out}(t) = \frac{1}{2}\sin\left(\frac{\pi}{V_\pi}V_{code}\right)s(t)V_x(t) + \frac{1}{4}\cos\left(\frac{\pi}{V_\pi}V_{code}\right)V_x^2(t) \\ - \frac{1}{2}\cos\left(\frac{\pi}{V_\pi}V_{code}\right) + \frac{1}{2}. \quad (4)$$

In Eq. (4), $V_{code}$ is not always $V_\pi/2$ because the waveform is not ideal. By comparing Eq. (3) with Eq. (4), it can be seen that in addition to introducing amplitude variations over time to the mixed term, the non-ideal PRBS also introduces additional second harmonic of the signal to be sampled with amplitude variations over time, as well as a term that varies over time and is independent of the signal to be sampled. Therefore, non-ideal PRBS will significantly affect the quality of the signal generated from the PD, resulting in a reduced reconstruction ability.

To reduce the impact of the non-ideal PRBS on microwave photonic RDs, the PRBS amplitude is reduced from $V_\pi/2$ in this work when the bias voltage is fixed to $V_\pi$. As discussed above, when PRBS switches between $V_\pi/2$ and $-V_\pi/2$, it can be considered that the actual bias point of the DD-MZM switches between the positive QTP and the negative QTP. Therefore, when the PRBS amplitude is reduced, as shown in Fig. 1(b), it can be considered that the actual bias point of the DD-MZM switches between two symmetric bias points lower than the QTP, that is low biasing the DD-MZM. It should be noted that although reducing the PRBS amplitude does not reduce the ratio of PRBS jitter to PRBS amplitude, the absolute value of the PRBS jitter is reduced. Returning to Eq. (4), with the decrease of the PRBS amplitude, although the overall amplitude of the mixed term is getting smaller and smaller, the jitter of the mixed term will also get smaller and smaller, and the mixed term will get closer and closer to the mixing result required by RD shown in Eq. (3). This is a beneficial change to improve the reconstruction ability. Of course, in this case, the aforementioned second harmonic term of the signal to be sampled and the term that changes with time and is independent of the signal to be sampled still exist, and their negative impact on the reconstruction ability of the RD increases with decreasing PRBS amplitude. This is a bad change to improve the reconstruction ability. In addition, as the bias point decreases with lower PRBS amplitude, shot noise and thermal noise will dominate, thereby reducing the SNR, which also harms the reconstruction ability [12, 13]. Therefore, appropriately lowering the bias point to ensure that the performance improvement brought by the compression of PRBS jitter exceeds the negative impact mentioned above can improve the reconstruction ability of the microwave photonic RD.

## 3. Experiment and results

A proof-of-concept experiment is implemented to verify the concept. First, to generate the non-ideal PRBS, an ideal 500-Mb/s PRBS in Matlab is filtered digitally to limit its bandwidth to 500 MHz, which is then uploaded to the AWG. The signal to be sampled is a multi-tone signal. The PRBS and the multi-tone signal both have a time length of 2 μs. The non-ideal PRBS and the multi-tone signal are mixed by the RD, and the output of the PD is sampled by the OSC at a sampling rate of 10 GSa/s to obtain 20,000 sampling points. The sampled signal is downsampled to 100 points after digital low-pass filtering, resulting in an equivalent sampling rate of only 50 MSa/s. Since the mixing quality in different positions of a PRBS code is different, we will select different starting points during downsampling in the subsequent experiment to study the results of sampling points falling in 20 different positions of a PRBS code. The reconstruction error is calculated by $\|x'-x\|_2^2 / \|x\|_2^2$, where $x'$ is the reconstructed signal, $x$ is the original signal.

A four-tone signal and a five-tone signal, which have a maximum frequency of 100 MHz, are sampled based on the proposed method. The reconstruction errors under different PRBS amplitudes are shown in Fig. 2. As can be seen from Fig. 2(a), when the PRBS amplitude changes from $0.432V_\pi$ to $0.177V_\pi$, the reconstruction error decreases at all sampling positions.

The errors of the 20 points in a PRBS amplitude are averaged to reflect the quality of signal recovery at this PRBS amplitude, and the results are shown in Fig. 2(b) with PRBS amplitude changed from $0.5V_\pi$ to $0.177V_\pi$. It can be seen that the average error is significantly reduced with the decrease of the PRBS amplitude, i.e., lowering the bias point of the DD-MZM. The corresponding results of the five-tone signal are shown in Fig. 2(c) and (d). As can be seen, in this case, the average error first decreases with the PRBS amplitude decreasing from $0.5V_\pi$ to $0.243V_\pi$, and then increases rapidly when the PRBS amplitude decreases further. This phenomenon is consistent with our previous analysis. In addition to the reduction of PRBS jitters, low biasing the DD-MZM will reduce the desired mixing term, make the unwanted second harmonic appear and increase, and in some cases worsen the SNR. The absence of a similar performance inflection point in the four-tone test, compared to the five-tone test, can be attributed to more second harmonics introduced by the five-tone signal. The feasibility of the proposed method and the correctness of our analysis are confirmed in Fig. 2.

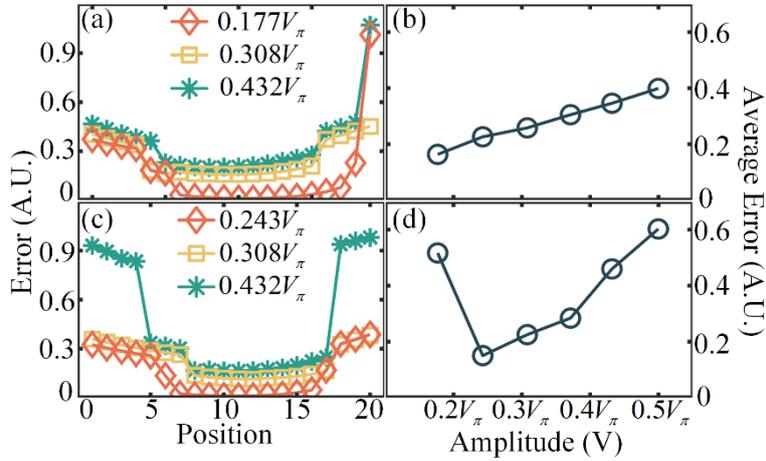

Fig. 2. Reconstruction errors at different positions of the PRBS code with different amplitudes for (a) four-tone signals and (c) five-tone signals. Average reconstruction errors for (b) four-tone signals and (d) five-tone signals under different PRBS amplitudes. The 500-Mb/s non-ideal PRBS has a bandwidth of 500 MHz.

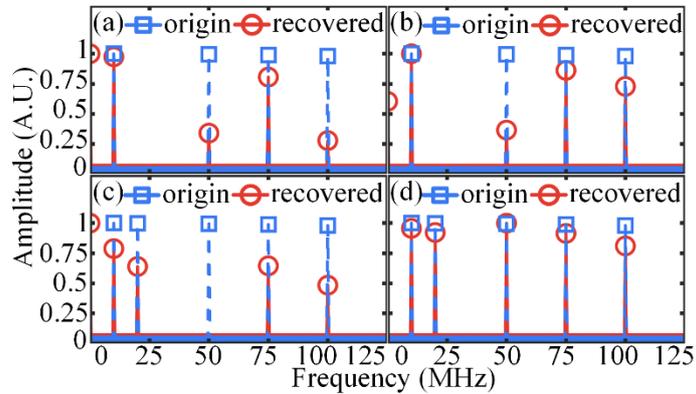

Fig. 3. Reconstruction results at position 16 of the PRBS code for a four-tone signal when the PRBS amplitude is (a) $0.5V_\pi$ and (b) $0.243V_\pi$ and for a five-tone signal when the PRBS amplitude is (c) $0.5V_\pi$ and (d) $0.243V_\pi$. The 500-Mb/s non-ideal PRBS has a bandwidth of 500 MHz.

To visualize the improvement of reconstruction quality via low biasing the DD-MZM, we further give the spectral comparison before and after reconstruction at position 16 when the PRBS amplitude is $0.5V_\pi$ and $0.243V_\pi$. As shown in Fig. 3(a) and (c), when the DD-MZM is equivalently biased at the QTPs as always done in the previous works, due to the non-ideality of the PRBS, although the frequencies of the multi-tone signals can be well restored, the spectrum amplitude of the reconstructed signal is quite different from that of the original signal. When the DD-MZM is low-biased at $0.243V_\pi$, as shown in Fig. 3(b) and (d), the reconstructed four-tone and five-tone signals are much improved compared with those in Fig. 3(a) and (c). The reconstruction error drops from 0.519 to 0.175 for the four-tone signal and from 0.498 to 0.03 for the five-tone signal. The DC component appears in Fig. 2(a) to (c), which is also caused by the limited reconstruction quality.

Then, the generality of the proposed method is further demonstrated by changing the non-ideality of the PRBS. The bandwidth limit for 500-Mb/s PRBS has been relaxed from 500 MHz in the previous experiment to 1 GHz. Indeed, the non-ideality of the PRBS is smaller in this case. Fig. 4 shows the reconstruction errors for the same four-tone and five-tone signals used above. Low-biasing the DD-MZM also greatly reduces the reconstruction error. The average reconstruction error in the 20 points reduced from 0.322 to 0.048 for the four-tone signal and from 0.342 to 0.052 for the five-tone signal. Compared with the results shown in Fig. 2, the results at different PRBS amplitudes are all improved, which is a reasonable result caused by the reduction of non-ideality. In addition, the inflection point is also observed in the five-tone test, and the inflection point does not appear in the four-tone test because we do not further reduce the PRBS amplitude. The reason why there is no further reduction in the PRBS amplitude in the experiment is that this is already the minimum amplitude after amplification when the AWG outputs the minimum signal amplitude.

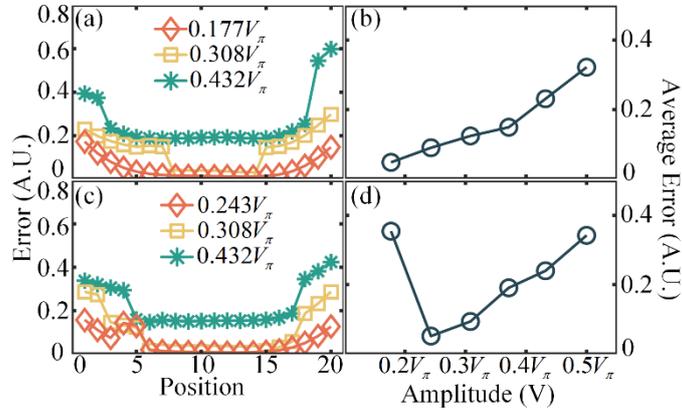

Fig. 4. (a) Reconstruction errors at different positions of the PRBS code with different amplitudes for (a) four-tone signals and (c) five-tone signals. Average reconstruction error for (b) four-tone signals and (d) five-tone signals under different PRBS amplitudes. The 500-Mb/s non-ideal PRBS has a bandwidth of 1 GHz.

Fig. 5 shows the spectra of the reconstructed signal and the original signal in different conditions. When the bias point is changed from the QTP to a lower bias point that has a PRBS amplitude of $0.243V_\pi$, the quality of signal reconstruction is greatly improved for both four-tone and five-tone signals.

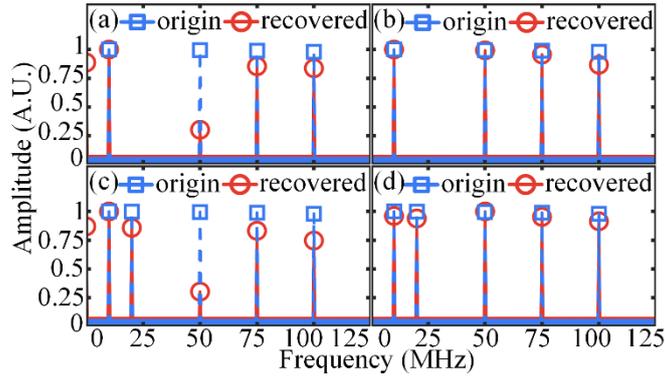

Fig. 5. Reconstruction results at position 16 of the PRBS code for a four-tone signal when the PRBS amplitude is (a)$0.5V_\pi$ and (b) $0.243V_\pi$ and for a five-tone signal when the PRBS amplitude is (c)$0.5V_\pi$ and (d) $0.243V_\pi$. The 500-Mb/s non-ideal PRBS has a bandwidth of 1 GHz.

## 4. Conclusion

In summary, a novel method for reducing the impact of non-ideal PRBS on microwave photonic RDs in a photonics-assisted CS system is proposed. To the best of our knowledge, this is the first method that has been experimentally validated to address the impact of non-ideal PRBS on signal reconstruction quality in microwave photonic RDs. By decreasing the amplitude of the non-ideal PRBS, we can equivalently establish a low bias point for the optical modulator in the microwave photonic RD. This adjustment is crucial for mitigating the impact arising from the random jitter of the non-ideal PRBS. The proposed method is of great significance for improving the performance and robustness of microwave photonic CS systems in practical applications.